\documentclass{Rinton-P9x6}
\newcommand{\postscript}[2]{\setlength{\epsfxsize}{#2\hsize}
   \centerline{\epsfbox{#1}}}
\usepackage{pifont}
\usepackage{amssymb}

\begin{document}

\title{Prospects for discovery of physics\\ beyond the Standard Model at\\
          the Pierre Auger Observatory}

\author{Luis A. Anchordoqui}

\address{Department of Physics, Northeastern University, Boston, MA 02115, USA
}


\maketitle

\abstracts{I summarize the discovery potential for physics beyond the electroweak
scale at the Pierre Auger Observatory.  This observatory is designed to
study ultra-high energy cosmic rays with unprecedented precision, with the
primary goal of shedding light on their composition and origins.
In addition, since the center-of-mass energies of Auger events
are well beyond those reached at terrestrial colliders, they provide
an opportunity to search for new physics. I discuss here some of the relevant
observables and techniques which may be used to weed out theories beyond
the standard model.}

\section{GZK--end of the cosmic ray spectrum?}

Shortly after the cosmic microwave background (CMB) was discovered, Greisen, Zatsepin, and Kuzmin (GZK) 
pointed out that the relic photons make the universe opaque to cosmic rays (CRs) of sufficiently high
energy.\cite{Greisen:1966jv} This occurs, for example, for protons with energies beyond the photopion 
production threshold, 
\begin{equation}
E_{p\gamma_{\rm CMB}}^{\rm th} = \frac{m_\pi \, (m_p + m_\pi/2)}{{\cal E}_{\rm CMB}} 
\approx 6.8 \times 10^{10}\,
\left(\frac{{\cal E}_{\rm CMB}}{10^{-3}~{\rm eV}}\right)^{-1} \,\,{\rm GeV}\,,
\label{1}
\end{equation}
where $m_p$ ($m_\pi$) denotes the proton (pion) mass and ${\cal E}_{\rm CMB} \sim 10^{-3}$~eV is a typical 
CMB photon energy. After pion production, the proton (or perhaps, instead, a neutron) emerges with at least 
50\% of the incoming energy. This implies that the nucleon energy changes by an $e$-folding after a 
propagation distance~$\lesssim (\sigma_{p\gamma}\,n_\gamma\,y)^{-1} \sim 15$~Mpc. Here, 
$n_\gamma \approx 410$~cm$^{-3}$ 
is the number density of the CMB photons, $\sigma_{p \gamma} > 0.1$~mb is the photopion 
production cross section, and $y$ is the average energy fraction (in the laboratory system) lost by a 
nucleon per interaction. For heavy nuclei, the giant dipole resonance can be 
excited at similar total energies and hence, for example, iron nuclei do not survive fragmentation 
over comparable distances. Additionally, the survival probability for extremely high energy ($\approx 10^{11}$~GeV)
$\gamma$-rays (propagating on magnetic fields~$\gg 10^{-11}$~G) to 
a distance $d$, \mbox{$p(>d) \approx \exp[-d/6.6~{\rm Mpc}]$}, becomes less than $10^{-4}$ after 
traversing a distance of 50~Mpc. All in all, as our horizon shrinks dramatically for energies
$\gtrsim 10^{11}$~GeV, one would expect a sudden cutoff in the energy spectrum if the CR sources follow a 
cosmological distribution.\cite{Bhattacharjee:1998qc} 

\begin{center}
\begin{figure}[tbp]
\begin{minipage}[t]{0.49\textwidth}
\postscript{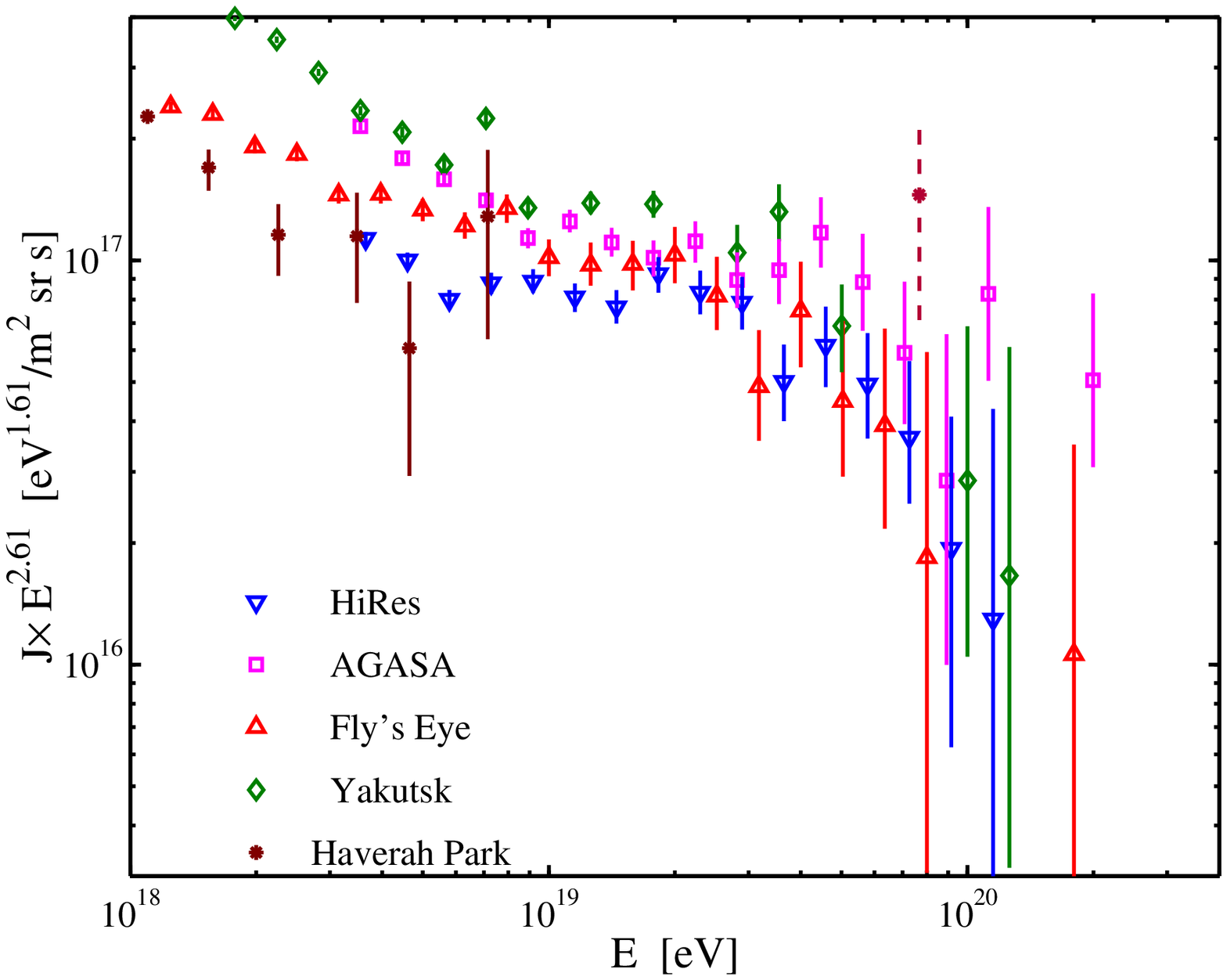}{1.0}
\end{minipage}
\hfill
\begin{minipage}[t]{0.49\textwidth}
\postscript{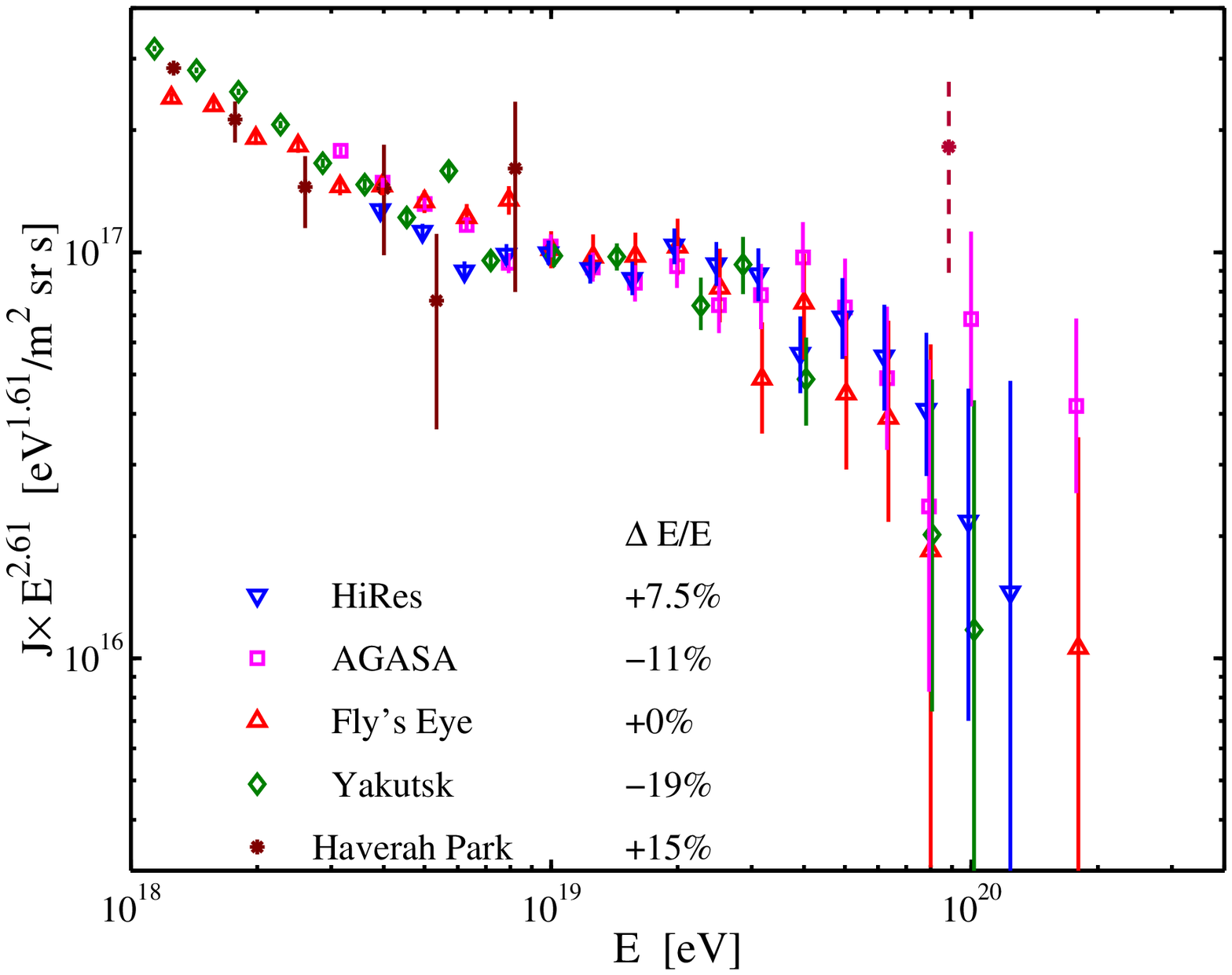}{1.0}
\end{minipage}
\caption{The left panel shows the upper end of the energy spectrum as reported by the 
different collaborations. The right panel shows the data after adjusting the absolute energy 
calibrations of the various experiments so as to bring the results from the different experiments 
into agreement at $10^{10}$~GeV. Here, the Fly's Eye energy is adopted as the standard. The fractional 
shifts in absolute energy scale, are well within the published systematic 
errors.\protect\cite{Bahcall:2002wi}}
\label{spectrum}
\end{figure}
\end{center}

At the beginning of summer 2002, in a pioneering paper Bahcall and 
Waxman\cite{Bahcall:2002wi} noted that 
the energy spectra of CRs reported by the AGASA, the Fly's Eye, the 
Haverah Park, the HiRes, and the Yakutsk Collaborations are consistent 
with the expected GZK cutoff. As one can readily see in Fig.~\ref{spectrum},
after small adjustments (within the known uncertainties) of the absolute 
energy scale, all these spectra are shown to be in agreement with each other 
for energies below $10^{11}$~GeV. A point worth noting at this juncture: the 
analysis that follows, which is based on counting events above roughly the 
expected cutoff, takes the data at face value and consequently does not 
attempt to evaluate possible correlated systematic errors. The particulars of 
the present sensitivity to a super-GZK flux (i.e., CR intensity 
beyond $10^{11}$~GeV) are given in Table 1.  Armed with the expected 
number of events and assuming Poisson statistics, it is easily seen 
that the existing data show significant evidence for a supression in the CR 
flux beyond $10^{11}$~GeV. Such a supression is found 
to be a $\sim 3.5\sigma$ effect according to the Fly's Eye normalization, 
increasing up to $\sim 8\sigma$ if the selected normalization is that of 
Yakutsk. It is important to emphasise that with data sets as small as these 
and with the inherent uncertainty associated with energy fluctuations, this 
result should be treated with great caution.

Systematic uncertainties in normalization have been thus far a headache common to all surface arrays.
Already in 1986, the SUGAR Collaboration instead of giving a unique primary energy for the observed 
events, they reported the showers' equivalent vertical muon number 
$N_{\mu{\rm v}}$ together with two possible conversions from 
$N_{\mu{\rm v}}$ to primary energy.\cite{Winn:un} Interestingly, if one adopts the Hillas conversion 
factor, for which the SUGAR integral energy spectrum is 
in good agreement with the AGASA spectrum at $10^{9}$~GeV,\cite{Bellido:2000tr} the number of observed 
super-GZK events  is consistent with the one expected from an extrapolation of the sub-GZK spectrum 
($\propto E^{-2.7}$) at the $1\sigma$ level. On the other hand, if one adopts the Sidney normalization 
there are no events with energy $> 10^{11}$~GeV. It should be noted that the SUGAR 
exposure given in Table 1 was obtained on the basis that the super-GZK detection probability over the 
entire array (with maximum collecting area of $70$~km$^{2}$) is $85\%$.\cite{Bell:gp} I have also 
assumed that the experiment operated in stable mode from January 1968 until February 1979, yielding a total 
area--time product of about 775~km$^2$ yr. However, since the sensitive area varied as the array was 
developed and as detectors require maintenance, the expected numbers of events given in the last row of 
Table 1 should be taken as upper limits. 

To make the situation more confusing, the calibrations from the various experiments are themselves 
moving targets.\cite{Ave:2001hq,Takeda:2002at,Abu-Zayyad:2002ta,Matthews:fc} Even in the time 
between this conference and the deadline for its proceedings the Yakutsk spectrum has change 
substantially.\cite{Ivanov:2003iv} Nowadays it seems easier to predict the stock market or even 
the weather in Boston than the next revision in the energy spectrum. 
{\it There is no doubt that more and better data is needed.}

The arrival directions of the super-GZK events have no apparent counterparts in the Galactic plane nor in 
the Local Supercluster. Furthermore, the data is consistent with an isotropic distribution of sources, in 
sharp contrast to the anisotropic distribution of light within 50 Mpc. The apparent isotropy in the arrival 
directions can be explained if the particle orbits are bent in extragalactic magnetic 
fields.\cite{Bhattacharjee:1998qc} However, the low statistics leaves a window open for less mundane 
explanations, which I discuss next. 

\begin{table}[htb]
\caption{Numbers of events observed with average energy $> 10^{11}$~GeV and incident zenith angle 
$< 45^\circ$. The super-GZK exposure of Volcano Ranch\protect\cite{Montanet},  
Haverah Park\protect\cite{Ave:2001hq}, Fly's Eye\protect\cite{Bird:wp}, 
Yakutsk\protect\cite{Bird:wp}, AGASA\protect\cite{Takeda:2002at}, 
HiRes\protect\cite{Abu-Zayyad:2002ta}, and SUGAR (see main text) is given in the $3^{\rm rd}$ column.  The 
last column indicates the expected number of events calculated based on the spectrum observed above 
$10^{10}$~GeV from Fly's Eye and Yakutsk, and assuming no change in the spectral index.}
\begin{center}
\footnotesize
\begin{tabular}{|c|c|c|@{}|cc|}
\hline 
Experiment & Events observed & Exposure [m$^2$ s sr] & \multicolumn{2}{@{} |c|}{Expected events}\\
    & & & ~~$_{\rm Fly's Eye}$ & $_{\rm Yakutsk}$ \\
\hline
Volcano Ranch & 1& $2.0 \times 10^{15}$ & 0.4 & 1.0 \\ 
Haverah Park & 0 &  $5.6 \times 10^{15}$& 1.2 & 2.9 \\
Fly's Eye & 1 & $2.6 \times 10^{16}$ & 5.4  & 13.4 \\
Yakutsk & 1  & $2.8 \times 10^{16}$ & 5.8 & 14.4 \\
AGASA & 11 & $5.1 \times 10^{16}$ & 10.6 & 26.2 \\
HiRes & 2 & $6.9 \times 10^{16}$ & 14.3 & 35.5 \\
$- - - - - - $&$- - - - - - -$&$- - - - - - -$&$- - -$&$- - -$\\
SUGAR & 5 & $< 3.1 \times 10^{16}$ & $<6.4$ & $< 15.9$ \\
\hline
\end{tabular}
\end{center}
\end{table}

\section{The usual suspects}

Physics from the most favored theories beyond the standard model (SM) like  
string/M theory, supersymmetry (SUSY), grand unified theories (GUTs), and TeV-scale gravity 
have been invoked to explain the super-GZK events. The conjectured origins fall into two 
basic categories: (i) exotic sources clustered nearby, (ii) GZK-evading messengers.  

The top suspect belongs to the first category. In this scenario charged and neutral primaries 
simply arise in the quantum mechanical decay of supermassive elementary $X$ 
particles. Sources of these exotic particles could be: (i) topological 
defects (TDs) left over from early universe phase transitions associated with 
the spontaneous symmetry breaking that underlies GUTs;\cite{Hill:1982iq} (ii)  
some long-lived metastable super-heavy relic (MSR) particles which may constitute
(a fraction of) the dark matter in galactic haloes. Arguably, the observed magnitude of the CMB 
fluctuations fixes the reheat temperature following inflation to $10^{13\pm1}$~GeV, allowing 
gravitational and thermal production of MSRs during the inflationary stage of the universe
with just the right mass ($m_X~>~10^{12}$~GeV)  for producing $10^{11}$~GeV secondary particles 
via decay.\cite{Berezinsky:1997hy} Due to their topological stability, the TDs (magnetic monopoles, 
cosmic strings, domain walls, etc.) can survive forever with $X$ particles 
($m_X \sim 10^{16} - 10^{19}$~GeV) trapped inside them. Nevertheless, from 
time to time, TDs can be destroyed through collapse, annihilation, or 
other processes, and the energy stored would be released in the form of 
massive quanta that would typically decay into quarks and leptons.
Discrete gauge symmetries or hidden sectors are introduced to stabilize
the MSRs and so generally higher dimensional operators are required 
to break the new symmetry 
super-softly to maintain an appreciable decay rate today (collisional 
annihilation has been considered, too). MSRs would also have 
quarks and leptons as the ultimate decay products. The strongly interacting quarks fragment 
into jets of hadrons containing mainly pions together with a 3\% admixture of nucleons. 
The injection spectrum is therefore a rather hard fragmentation-type shape with an upper limit usually 
fixed by the GUT scale. Of course, the precise decay modes of the $X$ particles and the detailed 
dynamics of the first secondary particles depend on the exact nature of the 
particles under consideration.\cite{Sigl:1998vz} However, one expects the bulk flow of outgoing 
particles to be almost independent of such details: all top-down scenarios predict a 
spectrum dominated by photons and neutrinos produced via pion decay.

The neutrino is the only known stable particle immune to the GZK degradation. 
The corresponding $\nu\bar{\nu}$ annihilation mean 
free path on the cosmic neutrino background, $\lambda_\nu
= (n_\nu \, \sigma_{\nu \bar{\nu}})^{-1} \approx  4
\times 10^{28}\,\,{\rm cm},$ is just above the present size of the horizon 
(recall that $H_0^{-1} \sim 10^{28}$~cm). One may then 
entertain the notion that neutrinos are indeed the super-GZK primaries. 
Unfortunately, perhaps, $\sigma_{\nu N}$ is, within the SM,  about five orders of magnitude too small 
to explain the observed atmospheric cascades. On the other hand, the limit imposed by 
unitarity is relatively weak and consequently does not impact on new interactions beyond the 
electroweak scale to increase significantly the neutrino-nucleon cross section.\cite{Domokos:1998ry} 

On a different track, if some flavor of neutrinos has a mass $\sim 0.1$~eV, 
the relic neutrino background
is a target for everyday weakly interacting neutrinos to form a 
$Z$-boson that subsequently decays producing a ``local'' flux of nucleons, 
photons and neutrinos.\cite{Weiler:1997sh}\footnote{Similarly, gravi-burst 
fragmentation jets can contribute to the super-GZK 
spectrum.\protect\cite{Davoudiasl:2000hv}} To reproduce 
the observed spectrum, the $Z$-burst mechanism requires very luminous 
sources of extremely high energy neutrinos throughout the 
universe (see Fig.~\ref{f2}).\cite{Kalashev:2001sh} 

A novel beyond--SM--model proposal to break the GZK barrier is to assume that 
ultrahigh energy CRs are not known particles but a new species of particle, 
generally referred to as the uhecron, $U$.\cite{Farrar:1996rg} The meager information we have 
about super-GZK particles allows a na\"{\i}ve description of the properties 
of the $U$. The muonic content in the atmospheric cascades suggests $U$'s should interact strongly. 
At the same time, if $U$'s are produced at cosmological distances, they 
must be stable, or at least remarkably long lived, with mean-lifetime 
$\tau \gtrsim 10^6 \, (m_U/3~{\rm GeV})\, (d/ {\rm Gpc})\,{\rm s},$ 
where $d$ is the distance to the source and $m_U$, the uhecron's mass.
Additionally, since the threshold energy increases linearly with $m_U$,
to avoid photopion production on the CMB $m_U \gtrsim 1.5$~GeV. In this 
direction, light  supersymmetric baryons (made from a light gluino + the 
usual quarks and gluons, $m_U \lesssim 3$~GeV) produce atmospheric cascades very similar to 
those initiated by protons.\cite{Berezinsky:2001fy}

Another interesting possibility in which super-GZK CRs can reach us from very distance 
sources may arise out of photons that mix with light axions in extragalactic magnetic 
fields.\cite{Csaki:2003ef} These axions would be sufficiently weakly coupled to travel large 
distances unhindered through space, 
and so they can convert back into high energy photons close to the Earth. An even 
more radical proposal postulates a tiny violation of local Lorentz invariance, 
such that 
some processes become kinematically 
forbidden.\cite{Coleman:1998en} In particular, photon-photon pair production and photopion 
production may be affected by Lorentz invariance violation. Hence, the absence of the 
GZK-cutoff would result from the fact that the threshold for photopion 
production ``disappears'' and the process becomes kinematically not allowed.

\section{Fingerprints in the sky: The distribution of arrival directions}

The distribution of arrival directions is perhaps the most helpful 
observable in yielding clues about cosmic ray origin. Neutral GZK-evading 
messengers should point back to distant active galaxies, thereby enabling 
point--source astronomy. The earliest super-GZK events did in fact point 
towards high-redshift compact, radio-loud quasars.\cite{Farrar:1998we} 
However, the current world data set show that such association is 
controversial.\cite{Sigl:2000sn}  Another revealing signature of discrete 
sources would be the clustering on a small angular scale. The data recorded 
by AGASA suggest that the 
pairing of events on the celestial sphere is occurring at a higher than 
random rate.\cite{Anchordoqui:2001qk} Moreover, event directions in a combined 
data sample from AGASA, Haverah Park, Yakutsk and Volcano Ranch 
also support no chance-association.\cite{Uchihori:1999gu} However, it is 
interesting to remark that to calculate a meaningful statistical 
significance of CR clustering, one must define the search 
procedure {\it a priori} in order to ensure it is not (inadvertently) 
devised especially to suite the
particular data set after having studied it. In the above mentioned analyses, 
for instance, the angular bin size was not defined ahead of time. 

Surface arrays in stable operation have nearly continuous observation over 
the entire year, yielding a uniform exposure in right ascension, $\alpha$. A traditional technique to 
search for large-scale anisotropies is then to fit the right ascension distribution of events to a 
sine wave with period $2\pi/m$ ($m^{\rm th}$ harmonic) to determine the components ($x, y$) of the Rayleigh 
vector\cite{Linsley}
\begin{equation}
x = \frac{2}{N} \sum_{i=1}^{N} \, \cos(m\, \alpha_i) \,, \qquad y = \frac{2}{N} \sum_{i=1}^{N} \,\, 
\, \sin( m\, \alpha_i)\,.
\end{equation}
The $m^{\rm th}$ harmonic amplitude of $N$ measurements $\alpha_i$ is given by the Rayleigh vector 
length ${\cal R}~=~(x^2~+~y^2)^{1/2}$. The expected length of such a vector for values randomly 
sampled from a uniform phase distribution is ${\cal R}_0~=~2/\sqrt{N}$.  The chance probability 
of obtaining an amplitude with length larger than that measured is
$p(\geq~{\cal R})~=~e^{-k_0},$ where $k_0~=~{\cal R}^2/{\cal R}_0^2.$ 
To give a specific example, a vector of length $k_0~\geq~6.6$ would be required to 
claim an observation whose probability of arising from random fluctuation
was 0.0013 (a ``$3\sigma$'' result). For the ultra-high energy~($>~10^{10.6}$~GeV) regime, all 
experiments\footnote{Since Fly's Eye has 
had a nonuniform exposure in sidereal time, ${\cal R}$ was computed using weighted 
showers.\cite{Cassiday:zh} A shower's weight depends on the hour of its sideral arrival time. 
The 24 different weights are such that every time bin has the same weighted number of showers.} to date
have reported $k_0 \ll 6.6$.\cite{Edge:rr} This does not imply an isotropic 
distribution, but it merely means that available data are too sparse to claim a statistically significant 
measurement of anisotropy by any of these experiments. In other words, there may exist anisotropies at 
a level too low to discern given existing statistics. For example, a clean signature of the MSR--$X$ 
hypothesis is the anisotropy imposed by the asymmetric position of the sun in the Galactic halo. 
As seen in the Northern hemisphere, the amplitude $\sim  0.3$ predicted by isothermal haloes with 
realistic core radii is in agreement with existing data (the most restrictive being the AGASA sample with 
${\cal R}_{m =1} \approx 0.3$ and $k_0 \approx 1.0$).\cite{Evans:2001rv} 

The $\alpha$ harmonic analyses are completely blind to intensity variations which depend only on 
declination, $\delta$.  Furthermore, combining anisotropy searches in $\alpha$ over a range of declinations 
could dilute the results, since significant but out of phase Rayleigh vectors from different 
declination bands can cancel each other out.  An unambiguous interpretation of anisotropy data requires 
two ingredients: {\it exposure to the full celestial sphere and analysis in terms of both celestial 
coordinates.} Though the statistics are very limited at present, the analysis of data from the SUGAR and 
the AGASA experiments, taken during a 10~yr period with nearly uniform exposure to the entire sky, shows 
no significant deviation from isotropy beyond $10^{10.6}$~GeV.\cite{Anchordoqui:2003bx}

\section{Smoking guns: Photon and neutrino fluxes}

Another telltale discriminator to be thought is the primary composition of the super-GZK events. Every 
model wherein the primaries arise from QCD jets (these models include 
$Z$-bursts, and decaying TDs and MSRs) produce many more mesons than baryons, and consequently the 
injection spectrum would be dominated by $\gamma$-rays produced through $\pi^0$ decay. Additionally, any 
technique used to distinguish photon-initiated showers from hadron-initiated showers would be well suited 
to test the validity of the photon-axion mixing model. 

At large zenith angles, hadrons and $\gamma$-rays develop cascades in the upper layers of the atmosphere.
The electromagnetic component, with mean interaction length $\sim 45-60$~g/cm$^2$, is absorbed long before 
reaching the ground by the greatly enhanced atmospheric slant depth ($\approx 3000$ g/cm$^2$ at $70^\circ$ 
from the zenith). Then surface arrays are practically only sensitive to the high energy muons created in 
the first few generations of particles. The shape of the shower front in this type of cascade is extremely 
flat (with  radius of curvature above $100$~km), and the particle time spread is very 
narrow ($\Delta t < 50$~ns). Since showers initiated by $\gamma$-rays produce fewer muons than those 
initiated by hadrons, one expects the 
rate of $\gamma$-ray showers detected by surface arrays to be reduced relative to the rate from hadron 
showers. Therefore, the determination of the CR-flux through vertical shower measurements using 
fluorescence eyes (which are fairly independent of the primary composition) provides a powerful tool for 
discriminating between hadron and $\gamma$-ray showers when comparing with the inclined shower rate. 
For example, a comparison of the showers recorded by the Haverah Park experiment (in the angular range 
$60^\circ< \theta < 80^\circ$) with predictions from the observed Fly's Eye spectrum
yields strong bounds on the $\gamma$-ray flux: above $10^{10}$~GeV, less than 48\% of the primary cosmic 
rays can be $\gamma$-rays and above $10^{10.6}$~GeV less than 50\% can be $\gamma$-rays. Both of these 
statements are made at the 95\%CL.\cite{Ave:2000nd} 

Even though $\gamma$-rays dominate at production, there are some viable TD scenarios which predict 
proton fluxes that are comparable or even higher than the $\gamma$-ray flux at all energies. Conversely,
due to the lack of absorption, relics clustered in the Galactic halo predict compositions directly 
given by the fragmentation function. Note that mechanisms which successfully deplete
the $\gamma$-rays (such as efficient absorption on the universal radio background) require an increase 
in the neutrino flux to maintain the overall normalization of the observed spectrum.
Within the SM, the mean free path of neutrinos is larger
than even the horizontal atmospheric depth. Neutrinos therefore
interact with roughly equal probability at any point in the atmosphere
and may initiate quasi-horizontal showers in the volume of air immediately above the
detector. These will appear as hadronic vertical showers, with large
electromagnetic components, curved fronts (a  radius of curvature of a few
km), and signals well spread over time (on the order of microseconds).

\begin{figure}
\begin{center}
\epsfig{file=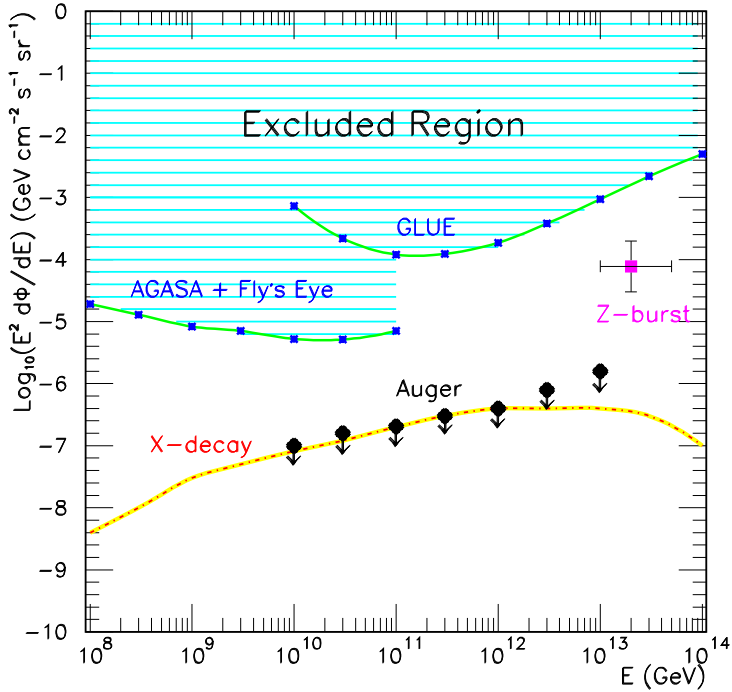, width=0.8\textwidth}
\end{center}
\caption{Current limits on neutrino fluxes. Upper bound 95\%CL
on \protect$d\Phi_{_{\langle \nu_i + \bar{\nu}_i \rangle}}/dE_{\nu_i}$
derived by replacing the combined exposures for deeply-developing showers 
of AGASA and Fly's Eye into Eq.~(\protect\ref{avg}).\protect\cite{Anchordoqui:2002vb} 
Upper bound 90\% CL on \protect$d\Phi_{_{\nu_\mu + \nu_e}}/dE_{\nu_i}$ from 
the non-observation of electromagnetic showers at GLUE induced by neutrinos 
interacting in the moon's rim.\protect\cite{Gorham:2001aj} The point with error
bars indicates the total neutrino flux required by the $Z$-burst mechanism\protect\cite{Kalashev:2001sh}
and the dotted line indicates a typical \protect$d\Phi_{_{\nu_\mu + \bar{\nu}_\mu}}/dE_{\nu_i}$ 
from top down cascades 
with  $m_{_X} = 10^{16}$~GeV.\protect\cite{Sigl:1998vz} In these models, 
\protect$d\Phi_{_{\nu_e + \bar{\nu}_e}}/dE_{\nu_i} \approx d\Phi_{_{\nu_\mu + \bar \nu_\mu}}/dE_{\nu_i}$. 
The diamonds indicate the sensitivity of the Pierre Auger Observatory, i.e., any flux lying above these  
points for at least one decade of energy will give more than 1 observed 
event per year.\protect\cite{Bertou:2001vm}} 
\label{f2}
\end{figure}

The event rate for quasi-horizontal deep showers from ultra-high
energy neutrinos is
\begin{equation}
N = \sum_{i,X} \int dE_i\, N_A \, \frac{d\Phi_i}{dE_i} \, \sigma_{i
N \to X} (E_i) \, {\cal E} (E_i)\ , \label{numevents}
\end{equation}
where the sum is over all neutrino species $i = \nu_e, \bar{\nu}_e,
\nu_{\mu}, \bar{\nu}_{\mu}, \nu_{\tau}, \bar{\nu}_{\tau}$, and all
final states $X$. $N_A = 6.022 \times 10^{23}$ is Avogadro's number,
and $d\Phi_i/dE_i$ is the source flux of neutrino species $i$, $\sigma$ as usual denotes 
the cross section, and ${\cal E}$ is the exposure measured in cm$^3$ w.e. sr time.
The Fly's Eye and the AGASA Collaborations have searched for quasi-horizontal showers that
are deeply-penetrating, with depth at shower maximum $X_{\rm max} > 2500$~g/cm$^2$. There 
is only 1 event that unambiguously passes this cut with 1.72 events expected from hadronic 
background, implying an upper bound of 3.5 events at 95\%CL from neutrino fluxes.\cite{Anchordoqui:2002vb}
Note that if the number of events integrated over
energy is bounded by 3.5, then it is certainly true bin by bin in
energy. Thus, using Eq.~(\ref{numevents}) one obtains
\begin{equation}
\sum_{i,X} \int_{\Delta} dE_i\, N_A \, \frac{d\Phi_i}{dE_i} \,
\sigma_{i N \to X} (E_i) \, {\cal E} (E_i)\  < 3.5\ ,
\label{bound}
\end{equation}
at 95\% CL for some interval $\Delta$. Here, the sum over $X$ takes
into account charge and neutral current processes.  In a logarithmic
interval $\Delta$ where a single power law approximation
\begin{equation}
\frac{d\Phi_i}{dE_i} \, \sigma_{i N \to X} (E_i) \, {\cal E} (E_i)
\sim E_i^{\alpha}
\end{equation}
is valid, a straightforward calculation shows that
\begin{equation}
\int_{\langle E\rangle e^{-\Delta/2}}^{\langle E\rangle e^{\Delta/2}}
\frac{dE_i}{E_i} \,
E_i\, \frac{d\Phi_i}{dE_i} \, \sigma_{i N \to X}  \, {\cal
E}  =   \langle \sigma_{i N\rightarrow X}\,
{\cal E} \, E_i\, d\Phi_i/dE_i \rangle \, \frac{\sinh \delta}{\delta}\, \Delta \,,
\label{sinsh}
\end{equation}
where $\delta=(\alpha+1)\Delta/2$ and $\langle A \rangle$ denotes the
quantity $A$ evaluated at the center of the logarithmic interval.  The
parameter $\alpha = 0.363 + \beta - \gamma$, where 0.363 is the
power law index of the SM neutrino cross
sections,\cite{Gandhi:1998ri} and $\beta$ and $-\gamma$ are the power
law indices (in the interval $\Delta$) of the exposure and flux
$d\Phi_i/dE_i$, respectively.  Since $\sinh \delta/\delta >1$, a
conservative bound may be obtained from Eqs.~(\ref{bound}) and
(\ref{sinsh}):
\begin{equation}
N_A\, \sum_{i,X} \langle \sigma_{i N\rightarrow X} (E_i)
\rangle \, \langle {\cal E} (E_i)\rangle\, \langle E_i
d\Phi_i/dE_i \rangle < 3.5/\Delta\ . \label{avg}
\end{equation}
By taking $\Delta =1$ as a likely interval in which the single power law behavior is
valid (this corresponds  to one $e$-folding of energy), and setting 
$\langle E_i d\Phi_i/dE_i \rangle = \frac{1}{6}
\langle E_\nu d\Phi_{\nu}/dE_\nu\rangle$ ($\Phi_{\nu} \equiv$ total neutrino flux) 
from Eq.~(\ref{avg}) it is straightforward to obtain 95\%CL upper limits on the
neutrino flux.\cite{Anchordoqui:2002vb} Similar concepts are  used by  the Goldstone 
Lunar Ultra-high energy neutrino Experiment (GLUE) to set upper bounds on the neutrino flux at the ultimate 
energy frontier.\cite{Gorham:2001aj} All these bounds are collected in Fig.~\ref{f2}.      

Unfortunately, due to the limited size of the current ultra-high energy CR sample, no 
conclusive statement can yet be made on the hypothetized models discussed in Sec.~2.
The jury awaits further evidence.

\section{Heavenly black holes: Probes of TeV-scale gravity}

It is intriguing --  and at the same time suggestive -- that the observed flux of CRs beyond the GZK-energy 
is well matched by the flux predicted for cosmogenic neutrinos. Of course, this is not a simple 
coincidence: any proton flux beyond
$E_{p\gamma_{\rm CMB}}^{\rm th}$ is degraded in energy by photoproducing $\pi^0$ and $\pi^\pm$, with the 
latter in turn decaying to produce cosmogenic neutrinos. The number of neutrinos produced in the 
GZK chain reaction compensates for their lesser energy, with the result that the cosmogenic flux matches 
well the observed CR flux 
beyond $10^{11}$~GeV. Recently, the prospect of an enhanced neutrino cross section has been explored in the 
context of theories with large compact dimensions. In these theories, the extra spatial dimensions are 
responsible for the extraordinary weakness of the gravitational force, or, in other words, the extreme 
size of the Planck mass. For example, if 
spacetime is taken as a direct product of a non-compact 4-dimensional manifold and a flat spatial 
$n$-torus $T^n$ (of common linear size $2\pi r_c$), one obtains a
definite representation of this picture in which the effective 4-dimensional 
Planck scale, $M_{\rm Pl} \sim 10^{19}$~GeV, is related to the fundamental scale of 
gravity, $M_D$, according to $M^2_{\rm Pl} = 8 \pi \, M_D^{2 +n} \, 
r_c^n$.\cite{Arkani-Hamed:1998rs} Within this framework,  virtual graviton exchange would disturb high 
energy neutrino interactions, and in principle, could increase the neutrino interaction cross section  
in the atmosphere by orders of magnitude beyond the SM value; namely 
$\sigma_{\nu N} \sim [E_\nu/(10^{10})~{\rm GeV}]$~mb.\cite{Domokos:1998ry} However, it is important to stress
that a cross section of $\sim 100$~mb would be necessary to approach obtaining consistency with observed 
showers which start within the first 50 g/cm$^2$ of the atmosphere. This is because Kaluza--Klein modes 
couple to neutral currents and the scattered neutrino carries away 90\% of the incident energy per 
interaction.\cite{Kachelriess:2000cb} Moreover, models which postulate strong neutrino interactions
at super-GZK energies also predict that moderately penetrating showers should be produced at lower energies,
where the neutrino-nucleon cross section reaches a sub-hadronic size. 
Within TeV scale gravity $\sigma_{\nu N}$ is likely to be sub-hadronic near the energy at which the 
cosmogenic neutrino flux peaks, and so moderately penetrating showers should be copiously produced.
Certainly, the absence of moderately penetrating showers in the CR data sample should be understood as a 
serious objection to the hypothesis of neutrino progenitors of the super-GZK events.\cite{Anchordoqui:2000uh}

\begin{center}
\begin{figure}[tbp]
\begin{minipage}[t]{0.49\textwidth}
\postscript{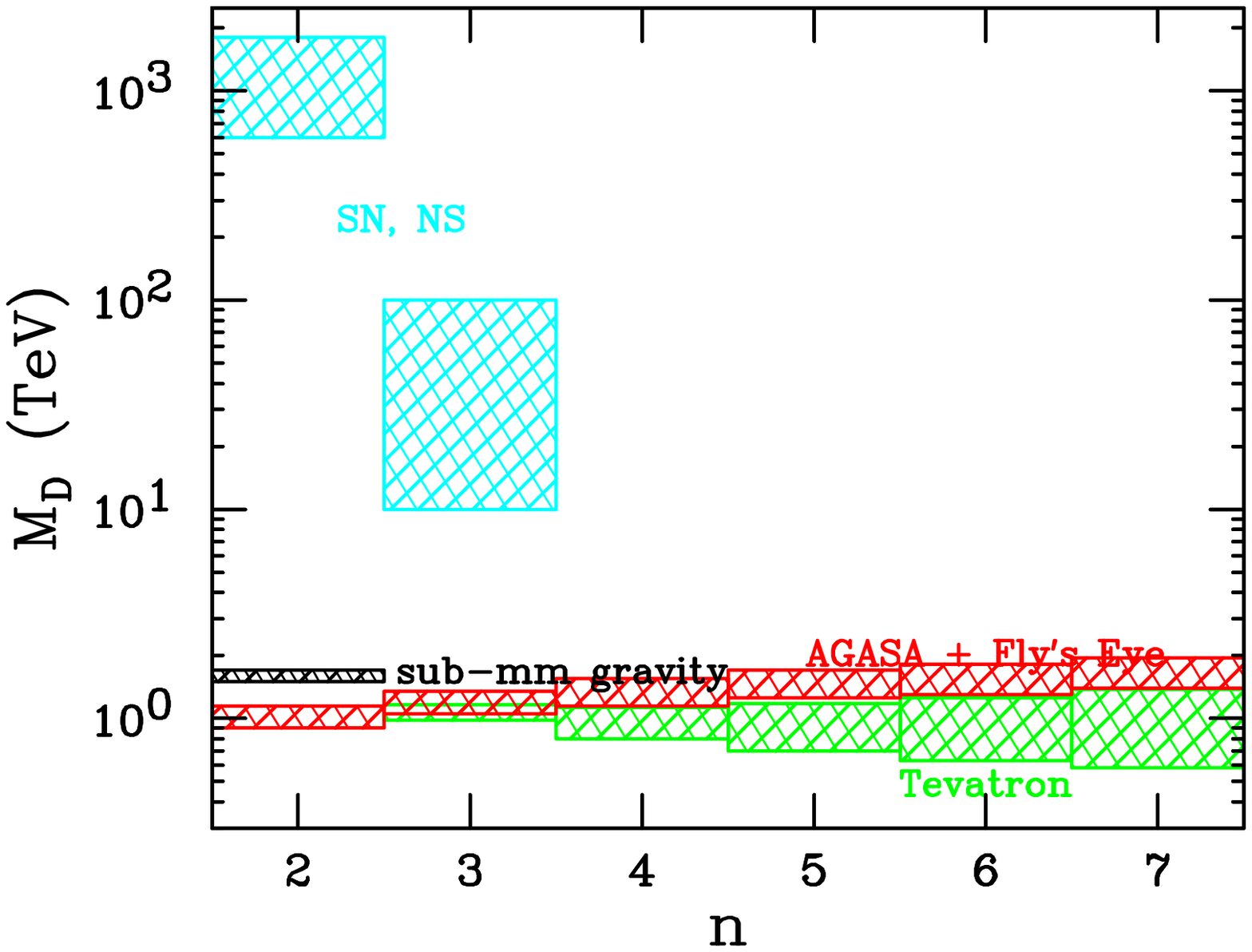}{1.0}
\end{minipage}
\hfill
\begin{minipage}[t]{0.49\textwidth}
\postscript{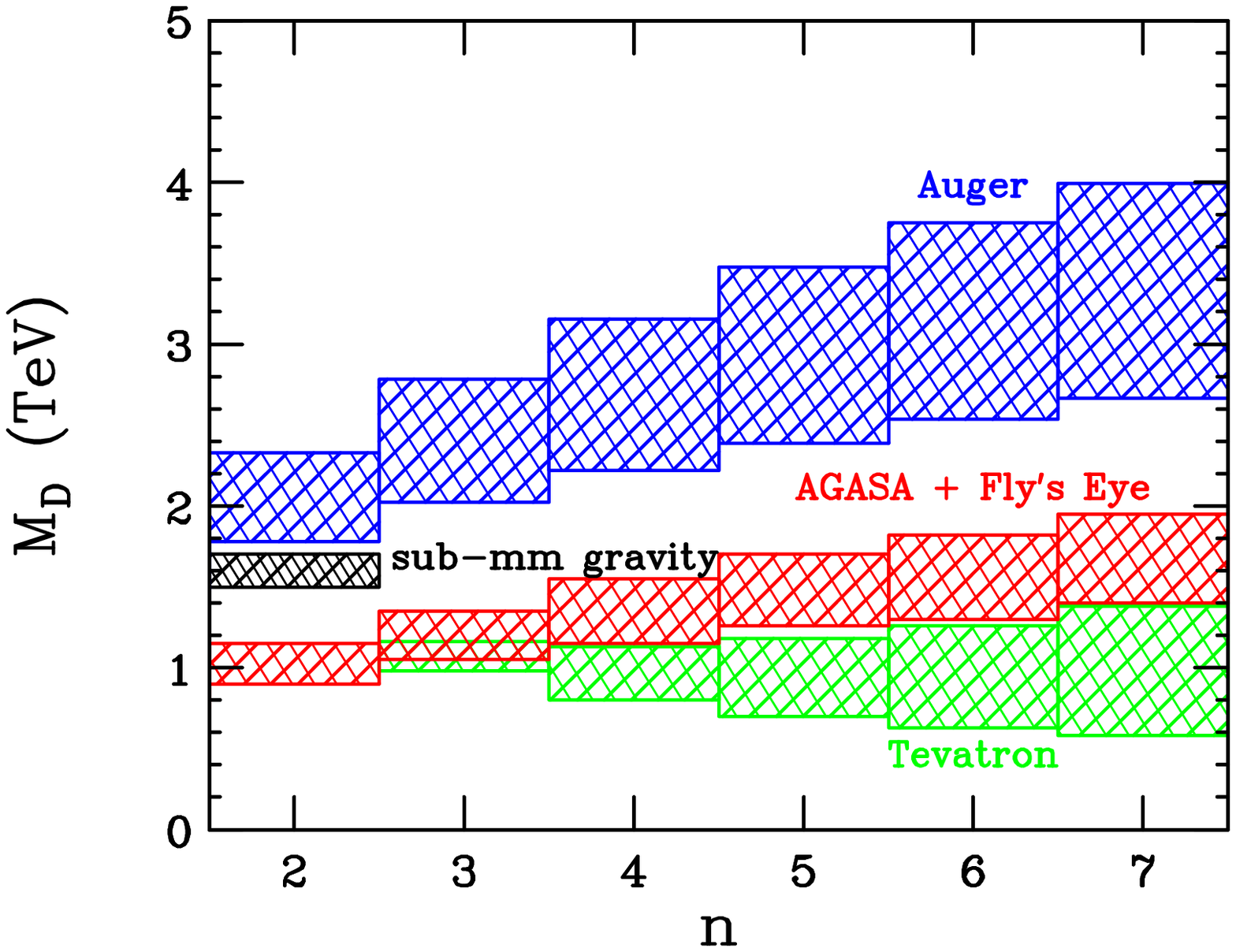}{1.0}
\end{minipage}
\caption{Left: Existing limits on the fundamental Planck scale $M_D$ from tests of
Newton's law on sub-millimeter scales,\protect\cite{Hoyle:2000cv} bounds on supernova cooling and
neutron star heating,\protect\cite{Cullen:1999hc} dielectron and diphoton production at the
Tevatron,\protect\cite{Abbott:2000zb} and non-observation of BH production at AGASA+Fly's 
Eye.\protect\cite{Anchordoqui:2002vb} Right: Comparison 
of current bounds on $M_D$ with future limits from the Southern Auger ground array, 
assuming 5 years of data and no excess above the SM neutrino background.\protect\cite{Anchordoqui:2001cg} 
The range in Tevatron bounds corresponds to the range of the brane softening parameter. The range in 
cosmic ray bounds is due to variations in the minimum BH mass.}
\label{bh}
\end{figure}
\end{center}

Large extra dimensions still may lead to significant increases in the neutrino cross section.
Should we be so lucky that this scenario is true, we might hope to observe 
black hole (BH) production (somewhat more masive than $M_D$) in elementary particle collisions with 
center-of-mass energies $\sqrt{s} \gtrsim {\rm TeV}$.\cite{Banks:1999gd} 
In particular, BHs occurring very deep in the atmosphere (revealed as intermediate states 
of ultrahigh energy neutrino interactions) could trigger quasi-horizontal 
showers and be detected by cosmic ray observatories.\cite{Feng:2001ib} Interestingly, 
$\sigma_{\nu N \rightarrow {\rm BH}} \propto M_D^{(-4+2n)/(1+n)}$. Therefore, using Eq.~(\ref{numevents}), 
the non-observation of the almost guaranteed flux of cosmogenic neutrinos can be translated into bounds 
on the fundamental Planck scale.\cite{Ringwald:2001vk,Anchordoqui:2001cg} In the case of $n$ extra spatial 
dimensions compactified on $T^n$ with a common radius, the bounds derived using AGASA+Fly's Eye exposure 
represent the best existing limits on the scale of TeV-gravity for $n\ge 4$ extra spatial
dimensions.\cite{Anchordoqui:2002vb}  A summary of the most stringent present bounds on $M_D$
for $n \ge 2$ extra dimensions is given in Fig.~\ref{bh}. Certainly, the lack of 
observed deeply-penetrating showers can be used to place more general, model-independent, bounds 
on $\sigma_{\nu N}$.\cite{Berezinsky:kz}

Up to now we have only discussed how to set bounds on physics beyond the
SM. An actual discovery of new physics in cosmic rays is a tall order
because of large uncertainties associated with the depth of the first 
interaction in the atmosphere, and the experimental challenges of reconstructing 
cosmic air showers from partial information. However, a similar technique to that employed
in discriminating between photon and hadron showers can be applied to search for
signatures of extra-dimensions. Specifically, if an anomalously large quasi-horizontal deep 
shower rate is found, it may be ascribed to either an enhancement of the incoming neutrino flux, or an 
enhancement in the neutrino-nucleon cross section. However, these two possibilities may be 
distinguished by separately binning events which arrive at very
small angles to the horizontal, the so-called ``Earth-skimming'' 
events.\cite{Feng:2001ue}  An enhanced flux will increase 
both quasi-horizontal and Earth-skimming event rates, whereas a large BH 
cross section suppresses the latter, because the hadronic decay products of 
BH evaporation do not escape the Earth's crust.\cite{Anchordoqui:2001cg}

\section{The PAO inquisition}

The Pierre Auger Observatory (PAO) is 
designed to work in hybrid mode, employing fluorescence detectors overlooking 
a ground array of deep water \v{C}erenkov radiators. 
During clear, dark nights,  events 
will be simultaneously observed by 
fluorescence light and particle detectors at ground level.
The PAO is expected to measure the energy, arrival direction and 
primary species with unprecedented statistical precision. It will eventually 
consist of two sites, one in the Northern hemisphere and one in the Southern, 
each covering an area of $3000$~km$^2$ and each consisting of 1600 particle detectors
overlooked by 4 fluorescence detectors. For showers with zenith angle $< 60^\circ,$ the 
overall acceptance (2 sites) is 14000~km$^2$ sr. The angular and energy resolutions of 
the ground array (without coincident fluorescence data) are typically less than $1.5^\circ$ and less than 
20\%, respectively. ``Golden events,'' those detected by
both methods simultaneously, will have a directional reconstruction 
resolution of about $0.3^\circ$ for energies near $10^{11}$~GeV. 
If an event trigger is 
assumed to require 5 detectors above threshold, the array is fully efficient 
at $10^{10}$~GeV. In three years of running, the surface arrays in both 
hemispheres will collect more than 1000 showers 
above $10^{10.6}$~GeV with approximately uniform sky 
exposure. This will enable a straightforward search for correlations with discrete 
sources and also a sensitive large scale anisotropy analysis.\cite{Sommers:2000us}

For showers with zenith angle exceeding $60^\circ,$ the aperture of PAO increases by
roughly 50\%. For a pure hadronic composition, there will be over 1000 well reconstructed 
events beyond $10^{10}$~GeV, with a mean energy error $\sim 25\%$. On 
the other hand, if $\gamma$-rays are dominant at high energy, the rate will be reduced by an order 
of magnitude allowing for a clear discrimination between these two cases. PAO will be able to 
establish strong bounds on the $\gamma$-ray flux at energies as high as $10^{11}$~GeV.\cite{Ave:2002ve}

In addition, PAO offers a window for neutrino astronomy 
above $10^{8}$~GeV. For standard neutrino interactions in the atmosphere, 
each site of PAO reaches $\sim 15$~km$^{3}$~w.e.~sr of target mass
around $10^{10}$~GeV, which is comparable to other 
neutrino detectors being planned.\cite{Capelle:1998zz} Moreover, the sensitivity of PAO
could be significantly enhanced by triggering on neutrinos that skim the 
Earth, traveling at low angles along chords with lengths of order their interaction 
length.\cite{Bertou:2001vm} As can be seen in Fig.~\ref{f2}, PAO will provide us with 
statistics to begin discriminating among the many promising ideas so far proposed to explain
the origin and nature of CRs above the GZK energy limits. Measurements of quasi-horizontal neutrino fluxes 
will also allow better limits to be placed on low scale gravity (see Fig.~\ref{bh}).

\section{Coda}

Statistics and better experimental handles should enable us to reconstruct the energy spectrum beyond 
$10^{11}$~GeV, to locate the CR sources in the sky, and to discern the primary chemical composition. 
Future CR data will not only provide clues to the CR origin, but could enhance our understanding of 
fundamental particle physics. For example, if CR primaries are found to have a significant $\gamma$-ray 
component above $10^{11}$~GeV, this could suggest an exotic ingredient in CRs,
such as the decay products of TDs/MSRs, and thus could provide insight into the description of the early 
universe as well as particle physics beyond the SM. The puzzle of ultrahigh energy CRs may even have 
something to say about issues as fundamental as local Lorentz invariance: the absence of photo-pion 
production above the GZK-limit would imply no cosmogenic neutrino flux and possibly undeflected pointing 
of the primary back to its source. Additionally, contrasting the observed quasi-horizontal  
neutrino flux with the expected neutrino flux can help to constrain TeV-scale gravity interactions
and improve current bounds on the fundamental Planck scale.
An optimist might even imagine the discovery of microscopic BHs, the telltale signature of the universe's 
unseen dimensions. We are entering  this new High Energy Physics era with the Pierre Auger Observatory.

\section*{Acknowledgments}

I am thankful to all my collaborators, especially to Jonathan Feng, Haim Goldberg 
and Al Shapere for enlightening discussions on neutrino air showers and TeV-scale gravity BHs, 
and to Steve Reucroft for a critical 
reading of this manuscript. I would also like to thank John Bahcall for valuable email 
correspondence and permission to reproduce Fig.~\ref{spectrum}. This work has been partially supported by 
the US National Science Foundation (NSF), under grant No. PHY-0140407.

\end{document}